\documentclass[twocolumn,showpacs,preprintnumbers,amsmath,amssymb]{revtex4}

\usepackage{amsmath,amssymb}
\usepackage[latin1]{inputenc}
\usepackage{dcolumn}
\usepackage{bm}

\newcommand{\diff}[0]{\text{d}}
\newcommand{\im}[0]{\text{Im}}
\newcommand{\re}[0]{\text{Re}}
\usepackage{graphicx}
\usepackage{dcolumn}
\usepackage{bm}

\begin{document}


\title{Negative refraction: A tutorial}

\author{Johannes Skaar}
\affiliation{Department of Electronics and Telecommunications, Norwegian University of Science and
Technology, NO-7491
Trondheim, Norway}
\affiliation{University Graduate Center, NO-2027 Kjeller, Norway}

\date{\today}

\begin{abstract}
The concepts of negative refractive index and near-field imaging are explained in the form of a tutorial.
\end{abstract}

\maketitle

\section{Tutorial}

It is possible to make artificial materials (metamaterials) with negative Re$\,\epsilon$ and Re$\,\mu$ in a limited bandwidth \cite{veselago,pendry1996,smith,pendry2004}. For microwave frequencies a negative Re$\,\mu$ can be achieved by creating arrays of metallic split-ring resonators, with dimensions much smaller than the wavelength. Negative Re$\,\epsilon$ is achieved by arrays of thin metallic wires simulating the response of a plasma. We limit the discussion to passive media. The refractive index of active media is discussed in detail in Refs. \cite{skaar06,skaar06b,nistad08}.

1. In a plasma the charges are free. Consider a free point charge $q$ in a uniform and monochromatic electric field $\mathbf E=E\exp(-i\omega t)\hat x$, where $\hat x$ is the unit vector in $x$ direction. (The physical electric field is given by the real part of $\mathbf E$.) Show that the displacement of the charge is
\begin{equation}
x=X\exp(-i\omega t),\quad X=-\frac{qE}{m\omega^2},
\end{equation}
where $m$ is the mass of the charge. If there are $N$ such free charges per unit volume, what is the polarization density associated with the charges? Argue that the relative permittivity can be written in the form
\begin{equation}
\label{relperm}
\epsilon_r=1-\frac{\omega_p^2}{\omega^2},
\end{equation}
and find $\omega_p$. Note that for $\omega<\omega_p$, $\epsilon_r$ is negative.

2. Use the Lorentz resonance model to explain that Re$\,\mu_r$ can be negative for frequencies above a strong magnetic resonance. Also explain that as the resonator strength approaches infinity, Im$\,\mu_r\rightarrow 0$ for the frequency where Re$\,\mu_r=-1$.

3. From now on, we assume that Im$\,\epsilon$ and Im$\,\mu$ are approximately zero so that $\epsilon$ and $\mu$ are real and negative. Furthermore, we assume that the medium is uniform. Using Maxwell's equations, show that for a plane wave, $({\bf E},{\bf H},{\bf k})$ form a left-handed set of vectors. Sketch the three vectors in addition to the Poynting vector ${\bf S}$.

4. Prove that the phase velocity $v\equiv\omega/k$ satisfies $v^2=(\omega/k)^2=1/\epsilon\mu$. Moreover, argue that the refractive index $n\equiv c/v$ satisfies $n^2={\epsilon_r\mu_r}$. Here $c$ is the vacuum light velocity. When $\epsilon_r$ and $\mu_r$ are positive, the solution is $n=+\sqrt{\epsilon_r\mu_r}$. However as we will see in the next question, when $\epsilon_r$ and $\mu_r$ are negative, we must choose $n=-\sqrt{\epsilon_r\mu_r}$.

5. A monochromatic, plane wave in vacuum is incident to a material with negative $\epsilon$ and $\mu$. The boundary is plane. Using the results in question 3 and the electromagnetic boundary conditions, explain that the angle of refraction must have opposite sign compared to the conventional case with positive $\epsilon$ and $\mu$. In other words, it makes sense to write Snell's law
\begin{equation}
\sin\theta_1=n\sin\theta_2,
\end{equation}
with $n=-\sqrt{\epsilon_r\mu_r}$, where $\theta_1$ and $\theta_2$ are the angles of incidence and refraction, respectively.

6. A monochromatic, plane wave in a medium with parameters $\epsilon_1$ and $\mu_1$ is incident to a medium with parameters $\epsilon_2$ and $\mu_2$. The wave vectors of the incident and transmitted waves are ${\mathbf k_1}=(0,k_{1y},k_{1z})$ and ${\mathbf k_2}=(0,k_{2y},k_{2z})$, respectively. The boundary is the plane $z=0$. Assume TE polarization, i.e., the electric field of the incident wave can be expressed ${\mathbf E}=(E,0,0)$. Argue that $k_{2y}=k_{1y}$. Using the electromagnetic boundary conditions, show that the reflection and transmission coefficients for the electric field can be written
\begin{subequations}
\label{fresnelgeneral}
\begin{align}
& r=\frac{\mu_2 k_{1z}-\mu_1 k_{2z}}{\mu_2 k_{1z}+\mu_1 k_{2z}},\\
& t=1+r.
\end{align}
\end{subequations}
Eqs. \eqref{fresnelgeneral} are general forms of the Fresnel equations, valid for complex $\epsilon_{1,2}$, $\mu_{1,2}$, $k_{1z}$, and $k_{2z}$. When $\epsilon$ and $\mu$ are real, and $\epsilon\mu$ is positive, deduce that the Fresnel equations can be written 
\begin{subequations}
\label{fresnelneg}
\begin{align}
& r=\frac{\eta_2\cos\theta_1-\eta_1\cos\theta_2}{\eta_2\cos\theta_1+\eta_1\cos\theta_2}\\
& t=1+r,
\end{align}
\end{subequations}
for propagating waves. Here $\theta_{1}$ and $\theta_2$ are the angle of incidence and refraction, respectively, and the wave impedances are given by $\eta_{1,2}=\sqrt{\mu_{1,2}/\epsilon_{1,2}}$, where the square root is taken to be positive. Conclude that the Fresnel equations give exactly the same reflection and transmission coefficient if the sign of $\epsilon$ and $\mu$ are flipped for one of the materials. 

\begin{figure}[t!!!]
\includegraphics[height=3cm,width=4cm]{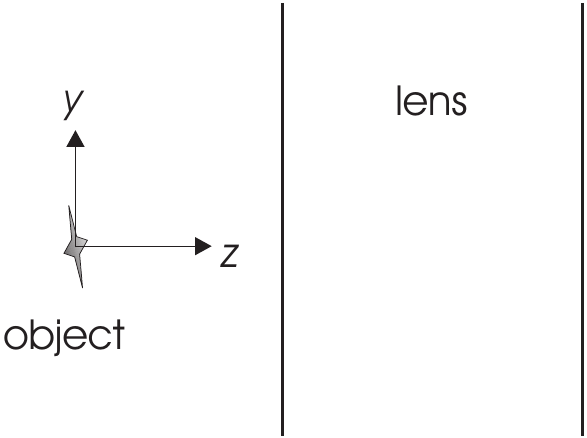}
\caption{The Veselago--Pendry lens.}
\label{fig:veselagolens}
\end{figure}
7. Consider a slab with $\epsilon_r=-1$ and $\mu_r=-1$, see Fig. \ref{fig:veselagolens}. Let the refractive index of the surrounding material be $1$. Trace the rays from the source, and demonstrate that the slab works as a lens (an image of the object is created on the right-hand side of the slab). Assume that the distance from the object to the lens is less than the thickness of the lens. For propagating waves, how much light is reflected at the boundaries?

8. The resolution limit when using a conventional lens is of the order of one wavelength. We will now demonstrate that the Veselago--Pendry lens, as described in the previous question, can provide unlimited resolution \cite{pendry2000}. Assume TE polarization so that the electric field can be written $\mathbf E=E\exp(-i\omega t)\hat x$. Furthermore, for simplicity we let the object be one-dimensional (that is, only dependent on $y$, not $x$), so that $E=E(y,z)$.

a) Assuming no reflections from the lens, explain that the total field can be written
\begin{equation}
\label{fourier}
E=\int_{-\infty}^{\infty} A(k_y)\exp(ik_yy+ik_zz)\diff k_y,
\end{equation}
where
\begin{equation}
\label{kvect}
k_z=\sqrt{\omega^2/c^2-k_y^2},
\end{equation}
for positive $z$ less than the distance from the object to the lens. Here $A(k_y)$ is the plane wave spectrum of the field at the object ($z=0$). Using \eqref{fourier} and \eqref{kvect}, explain that a conventional lens cannot resolve details in the object that is smaller than one wavelength.

b) Use one of Maxwell's equations (Faraday's law) to show that $(\partial E/\partial z)/\mu$ must be continuous at the boundaries. Hint: Use the boundary condition for the magnetic field.

c) We have so far argued that propagating waves do not reflect at the boundaries. Assuming that this is also the case for evanescent waves, and using the result of 8b), show that $k_z$ must change sign at the boundaries. Explain that the evanescent fields are amplified in the lens so that the resolution of the image has no limit.

\section{Solution}

1. Using Newton's law in the frequency domain, we obtain $qE=-m\omega^2 X$. The polarization density (in the frequency domain) becomes $P=NqX=-Nq^2E/m\omega^2$, giving the susceptibility $\chi=-Nq^2/m\epsilon_0\omega^2$. The permittivity is therefore given by \eqref{relperm}, with $\omega_p^2=Nq^2/m\epsilon_0$.

\begin{figure}[h!!!]
\includegraphics[height=4.5cm,width=5.5cm]{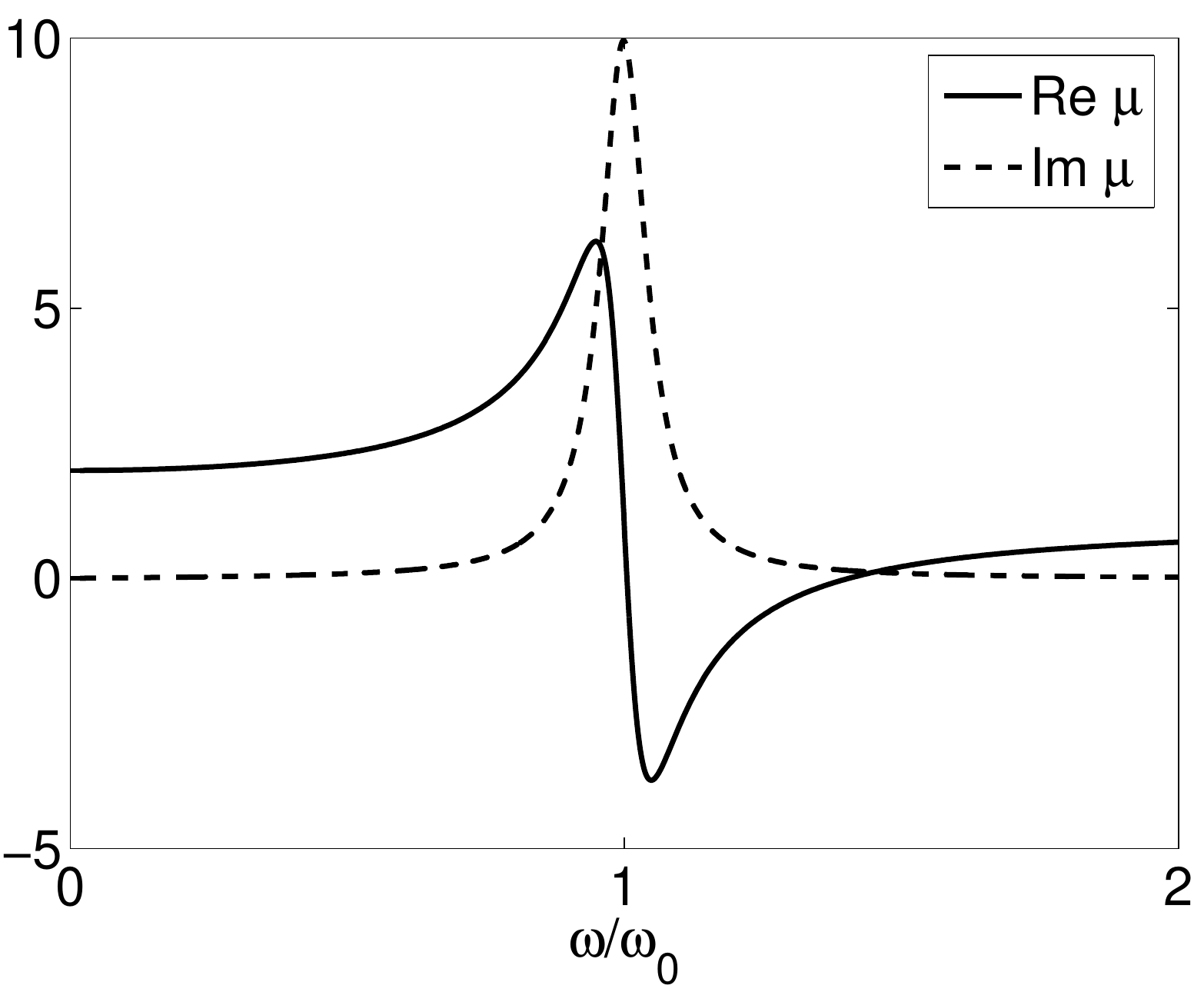}
\caption{A Lorentzian resonance of width $\Delta\omega=0.1\omega_0$.}
\label{fig:lorentz}
\end{figure}

2. A Lorentzian resonance is a damped harmonic oscillator model with permeability of the form 
\begin{equation}
\mu_r=1+\chi_0\frac{\omega_0^2}{\omega_0^2-\omega^2-i\omega\Delta\omega},
\end{equation}
or
\begin{subequations}
\label{lorentz}
\begin{align}
&\re\,\mu_r=1+\chi_0\frac{\omega_0^2(\omega_0^2-\omega^2)}{(\omega_0^2-\omega^2)^2+(\omega\Delta\omega)^2},\\
&\im\,\mu_r=\chi_0\frac{\omega_0^2\omega\Delta\omega}{(\omega_0^2-\omega^2)^2+(\omega\Delta\omega)^2}.
\end{align}
\end{subequations}
Here $\omega_0$ and $\Delta\omega$ are the central frequency and bandwidth, respectively, and the resonator strength is described by the constant $\chi_0$. Above the central frequency, the real part is negative provided the resonance is sufficiently strong. The permeability of a relatively strong resonance is plotted in Fig. \ref{fig:lorentz}. Note that when $\omega$ increases, $\im\,\mu_r$ approaches zero faster than does $\re\,\mu_r-1$. This can also be realized directly from the expressions \eqref{lorentz}. Hence, when $\chi_0\rightarrow\infty$, $\im\,\mu_r$ approaches zero for the frequency where $\re\,\mu_r=-1$.

\begin{figure}[t!!!]
\includegraphics[height=3.5cm,width=4.5cm]{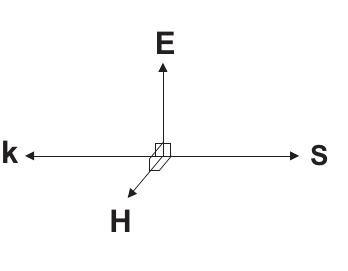}
\caption{The field vectors of a plane wave.}
\label{fig:vectors}
\end{figure}

3. We are searching for plane wave solutions to Maxwell's equations, i.e., solutions on the form ${\bf E}={\bf E_0}\exp(i{\bf k}\cdot{\bf r}-i\omega t)$ and ${\bf H}={\bf H_0}\exp(i{\bf k}\cdot{\bf r}-i\omega t)$. Substituting into the two curl equations, we find 
\begin{subequations}
\label{kEkH}
\begin{align}
&{\bf k}\times{\bf E}=\omega\mu{\bf H},\label{kE}\\
&{\bf k}\times{\bf H}=-\omega\epsilon{\bf E}\label{kH}.
\end{align}
\end{subequations}
Since $\epsilon$ and $\mu$ are negative, it follows that $({\bf E},{\bf H},{\bf k})$ form a left-handed set, see Fig. \ref{fig:vectors}. For this reason, such media are sometimes referred to as ``left-handed media''.

4. Writing ${\bf E}=E\exp(-i\omega t)\hat x$, ${\bf H}=H\exp(-i\omega t)\hat y$, and ${\bf k}=-k\hat z$, \eqref{kEkH} becomes
\begin{subequations}
\label{kEkHscalar}
\begin{align}
&kE=-\omega\mu H,\\
&kH=-\omega\epsilon E.
\end{align}
\end{subequations}
Multiplying the two equations \eqref{kEkHscalar}, we obtain $k^2=\mu\epsilon\omega^2$ or $(\omega/k)^2=1/\mu\epsilon$. Thus the refractive index satisfies $n^2=c_0^2/(\omega/k)^2=\mu_r\epsilon_r$.  

\begin{figure}[t!!!]
\includegraphics[height=4cm]{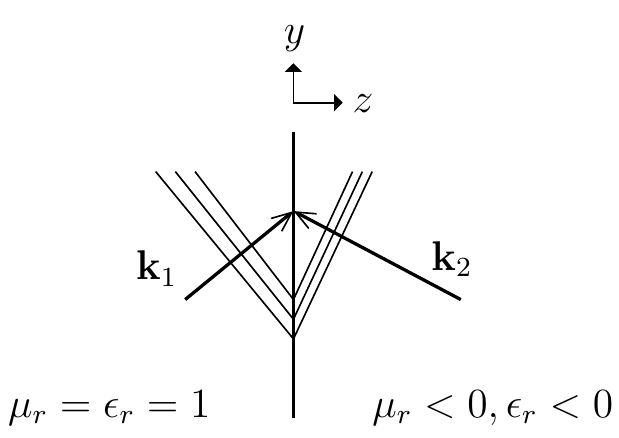}
\caption{Snell's law. The wave vectors in the left and right medium are denoted ${\bf k_1}$ and ${\bf k_2}$, respectively. Wavefronts associated with the incident and the refracted waves are shown as lines orthogonal to the wave vectors. Since the tangential fields are continuous, the wave fronts projected on the boundary must propagate in the same direction on each side. In this particular case the material on the right-hand side has $\epsilon_r\mu_r>1$.}
\label{fig:snell}
\end{figure}

5. First we note that the normal component of the Poynting vector must be continuous since there is no accumulation of energy at the boundary. Thus, the wave vector ${\bf k}_2$ has a negative $z$ component, see Fig. \ref{fig:snell}. On the other hand, the boundary conditions require that the tangential components of the electric and magnetic fields are continuous. This implies phase match across the boundary. In other words, the incident wave projected on the boundary must propagate in the same direction as the projection from the refracted beam. Putting these results together, we find negative refraction, see Fig. \ref{fig:snell}.

6. The phase match at the boundary (see above) requires that $k_{1y}=k_{2y}$. At $z=0$, the incident, reflected, and transmitted field amplitudes are $E$, $rE$ and $tE$, respectively. Imposing continuity of the tangential electrical field yields
\begin{equation}
\label{contE}
1+r=t.
\end{equation}
The magnetic field associated with the incident wave is found using \eqref{kE}: ${\mathbf H}=(0,Ek_{1z}/\omega\mu_1,-Ek_{1y}/\omega\mu_1)$. Similarly, the reflected and transmitted magnetic fields are $(0,rE(-k_{1z})/\omega\mu_1,-rEk_{1y}/\omega\mu_1)$ and $(0,tEk_{2z}/\omega\mu_2,-tEk_{2y}/\omega\mu_2)$, respectively. Hence, continuity of the tangential magnetic field gives
\begin{equation}
\label{contH}
k_{1z}/\mu_1-rk_{1z}/\mu_1=tk_{2z}/\mu_2.
\end{equation}
Combining \eqref{contE} and \eqref{contH} leads to the desired result \eqref{fresnelgeneral}. For $\epsilon_{1,2}$ and $\mu_{1,2}$ positive, $k_{1z}$ and $k_{2z}$ are positive as well (assuming propagating waves), and \eqref{fresnelneg} follows using $\cos\theta_{1}=k_{1z}/k_1$ and $\cos\theta_{2}=k_{2z}/k_2$. If, for example, $\epsilon_{2}$ and $\mu_{2}$ change sign, the reflection coefficient do not change since both $\mu_2$ and $k_{2z}$ change sign.  

\begin{figure}[t!!!]
\includegraphics[height=3cm,width=5.7cm]{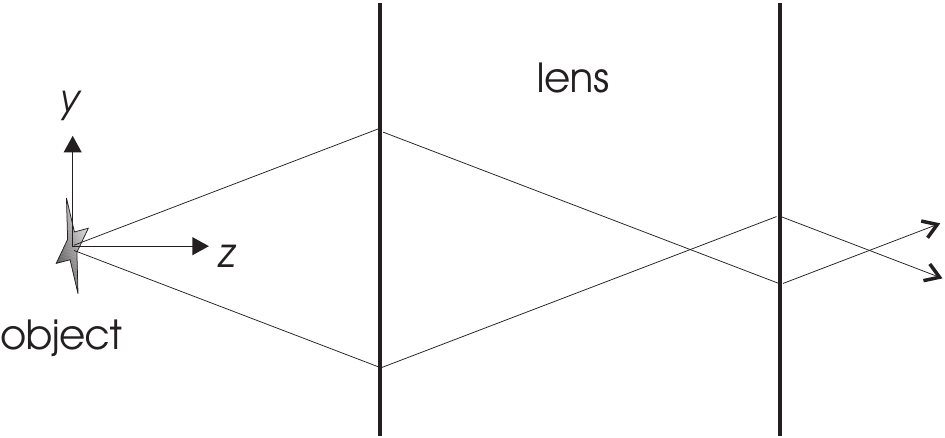}
\caption{Ray tracing from the object to the image. The refractive index of the lens material is $n=-1$.}
\label{fig:raytrace}
\end{figure}

7. By tracing the rays we conclude that there is an image inside the lens, and also at the right-hand side, see Fig. \ref{fig:raytrace}. There is no reflection at the boundaries according to the Fresnel equations.

8a) Taking a Fourier transform of the source field, we obtain a spectrum $A(k_y)$. Eq. \eqref{fourier} then follows as the inverse transform in the special case $z=0$. For $z>0$ the field is propagated from $z=0$ to $z$ using Maxwell's equations. This is achieved for each plane wave component by the transformation $A(k_y)\exp(ik_yy)\rightarrow A(k_y)\exp(ik_yy+ik_zz)$, where $k_z^2=(\omega/c)^2-k_y^2$. Note that the plane wave components associated with large spatial frequencies in the object are evanescent (i.e., $k_z$ is imaginary). The largest $k_y$ that corresponds to a propagating wave is $k_{y,\max}=\omega/c$. A conventional lens is not able to restore the evanescent field, so the finest details in the image is $\Delta y=2\pi/k_{y,\max}=2\pi/(\omega/c)=\lambda$, where $\lambda$ is the vacuum wavelength.

8b) Faraday's law $\nabla\times{\bf E}=i\omega\mu{\bf H}$ becomes in this case $\hat y \partial E/\partial z - \hat z\partial E/\partial y=i\omega\mu{\bf H}$. The tangential component of the magnetic field is the $y$ component, and consequently $(\partial E/\partial z)/\mu$ must be continuous at the boundary.

8c) Since $\mu$ changes sign at each boundary, so does $\partial E/\partial z$. Assuming no reflections, the field for each $k_y$ has only a single plane wave component $A(k_y)\exp(ik_yy+ik_zz)$. This means that $k_z$ changes sign at each boundary. Thus, an evanescent decaying wave in vacuum results in a growing wave in the lens material. From the symmetry in Fig. \ref{fig:raytrace}, it is clear that the wave propagates an equal distance along the $z$ axis in both materials, so the resulting decay/amplification is 1. In other words, the evanescent field is perfectly reconstructed in the image so that the resolution has no limit.

Note that the evanescent field amplification is infinite in the limit of infinite spatial frequencies or in the limit of infinite slab thickness. Nevertheless the evanescent fields do not transport energy in the $z$ direction as can readily be verified by computing the complex Poynting vector. 

\appendix
\section*{Detailed analysis of the perfect lens}
The argument in 8c), that $k_z$ changes sign at the boundary for evanescent waves, is not rigorous since we have not proved that the evanescent waves do not reflect at the boundaries. Rigorously, the fields must be found by solving Maxwell's equations for arbitrary $\epsilon$ and $\mu$, where $\im\,\epsilon>0$ and $\im\,\mu>0$ \cite{landau_lifshitz_edcm} for passive media. As opposed to the analysis above, we now consider TM polarization, i.e., the magnetic field points in the $x$-direction. We consider each spatial frequency $k_y$ separately, and define the transmission coefficient $T$ as the ratio between the plane wave amplitudes at the image and at the object. In each of the three regions (to the left of the lens, in the lens, and to the right of the lens), the magnetic field can be written in the form $A^+\exp(ik_y y+ik_z z)+A^-\exp(ik_y y-ik_z z)$. Here $k_z^2=\omega^2/c^2-k_y^2$ outside the lens, while $k_z^{2}=\epsilon_r\mu_r\omega^2/c^2-k_y^2$ in the lens (let us call this latter wavenumber $k_z'$). The constants $A^+$ and $A^-$ in each region (that is, 6 constants) are found as follows. The $A^-$ constant to the right of the lens is zero since nothing is sent from the other side. The $A^+$ constant to the left of the lens can be normalized to unity. The remaining 4 constants are found by matching the tangential electric and magnetic fields at both surfaces. The transmission coefficient $T$ is identified from $A^+$ to the right of the lens, at $z=2d$, where $d$ is the thickness of the lens. The result is \cite{ramakrishna2002,johansen}
\begin{equation}\label{tc1}
T=\frac{4\epsilon_r k_z k'_z e^{i(k_z-k'_z)d}}{(\epsilon_r k_z+k'_z)^2
e^{-2ik'_z d}-(\epsilon_r k_z-k'_z)^2}.
\end{equation}
Here the sign of $k_z$ must be
chosen such that $\im\, k_z\geq 0$; otherwise the field would be infinite for $z=\infty$. Remarkably, the sign of $k_z'$ does not matter in \eqref{tc1}. Taking the limit $\epsilon_r\to -1$ and $\mu_r\to -1$ we find $|T|\to 1$. In other words, both propagating and evanescent waves are perfectly reconstructed at the image plane.

\begin{figure}[]
\includegraphics[height=5.5cm]{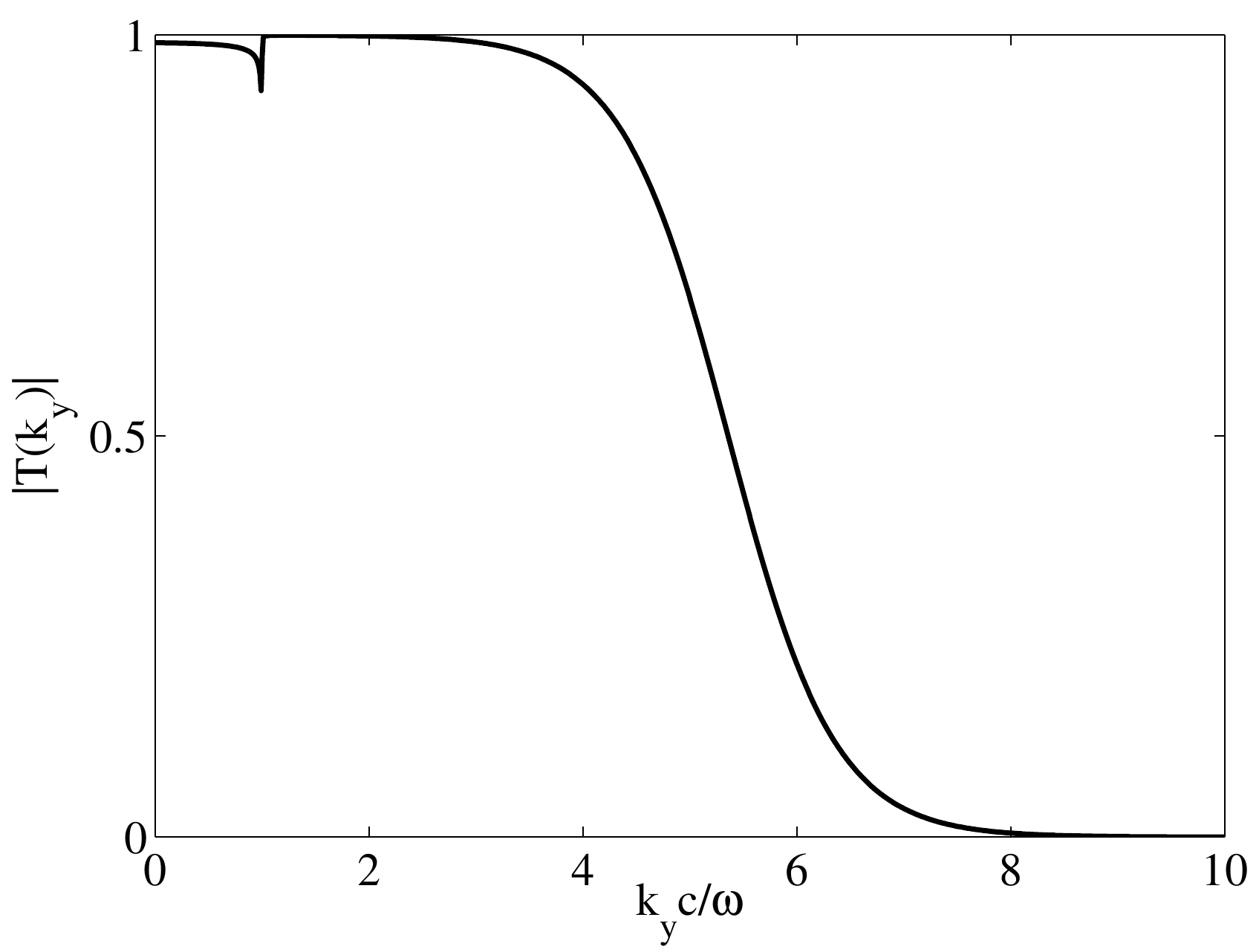}
\caption{The absolute value of the transmission coefficient $|T|$ as a function of spatial frequency $k_y$. The parameters are $\epsilon_r=\mu_r=-1+0.01i$ and $\omega d/c=1$. There is a small distortion of the propagating components (with $k_y\lesssim \omega/c$). The transmission is larger than 0.5 up to the value predicted by \eqref{resolution1}, $k_{y\max}=5.3\omega/c$.}
\label{fig:T}
\end{figure}

In a passive medium, there will always be losses. As an example of the transmission coefficient in a realistic medium, see Fig. \ref{fig:T}. To facilitate the analysis of such a realistic case, we limit the discussion to large values of $k_y$, corresponding to near-fields that decay
exponentially in vacuum. We therefore assume that $k_y \gg\omega/c$.
With the additional assumptions that $|\epsilon_r+1|\ll 1$, $|\mu_r|$ at
most of the order $1$, and $d$ at most of the order a vacuum
wavelength ($\omega d/c\lesssim 1$), we obtain
\begin{equation}\label{tc2}
\frac{1}{T}\approx 1-\frac{\left[(1-\epsilon_r^2)-(\epsilon_r-\mu_r)
(\omega/ck_y)^2\right]^2}{16}\exp(2k_y d),
\end{equation}
where we have used that 
\begin{align}
\epsilon_r k_z+k'_z &= \frac{\epsilon_r^2
k_z^2-k_z'^{2}}{(\epsilon_r k_z-k'_z} \\
&=\frac{(1-\epsilon_r^2)k_y^2+\epsilon_r(\epsilon_r-\mu_r)(\omega/c)^2}{\epsilon_r
k_z-k'_z}. \nonumber
\end{align}

We take the resolution $k_{y,\max}$ to be the largest value of
$k_y$ such that the modulus of the second term on the right-hand side of \eqref{tc2} is equal to 1. This definition makes sense no matter what this term's phase happens to be, since the exponential factor $\exp(2k_yd)$ will force $|T|$ to decrease rapidly when $k_y$ gets larger than $k_{y,\max}$. In general, the resolution becomes a nontrivial function of $\epsilon_r$ and $\mu_r$, but if $|\epsilon_r-\mu_r|(\omega/ck_y)^2\ll 2|\epsilon_r+1|$, then
\begin{equation}\label{resolution1}
k_{y,\max} = - \frac{1}{d}\ln{\frac{|\epsilon_r+1|}{2}}.
\end{equation}
We note that to obtain a linear increase in resolution, the losses must be exponentially small!

By the assumption $\omega/ck_y\ll 1$, the requirement
$|\epsilon_r-\mu_r|(\omega/ck_y)^2\ll 2|\epsilon_r+1|$ can be rewritten as
\begin{equation}\nonumber
|\epsilon_r-\mu_r|(\omega/c)d\lesssim -2|\epsilon_r+1|\ln\frac{|\epsilon_r+1|}{2}.
\end{equation}
Thus \eqref{resolution1} is always valid if $\epsilon_r=\mu_r$. Also, it
is valid provided the lens is sufficiently thin. Note that in the
latter case the resolution is independent of $\mu_r$. If we had chosen
the opposite polarization (i.e., electric field along the $x$-axis),
we would have arrived at exactly the same result only with the roles
of $\epsilon_r$ and $\mu_r$ interchanged. In other words, for a
one-dimensional object it is sufficient to have one of the
parameters $\epsilon_r$ and $\mu_r$ close to $-1$ with a small imaginary
part \cite{pendry2000}. For a two-dimensional object, both
polarizations are necessarily present; thus both $\epsilon_r$ and
$\mu_r$ should be close to $-1$.
 
\bibliography{causbib,nbib}

\def\cprime{$'$}
\begin{thebibliography}{11}
\expandafter\ifx\csname natexlab\endcsname\relax\def\natexlab#1{#1}\fi
\expandafter\ifx\csname bibnamefont\endcsname\relax
  \def\bibnamefont#1{#1}\fi
\expandafter\ifx\csname bibfnamefont\endcsname\relax
  \def\bibfnamefont#1{#1}\fi
\expandafter\ifx\csname citenamefont\endcsname\relax
  \def\citenamefont#1{#1}\fi
\expandafter\ifx\csname url\endcsname\relax
  \def\url#1{\texttt{#1}}\fi
\expandafter\ifx\csname urlprefix\endcsname\relax\def\urlprefix{URL }\fi
\providecommand{\bibinfo}[2]{#2}
\providecommand{\eprint}[2][]{\url{#2}}

\bibitem[{\citenamefont{Veselago}(1968)}]{veselago}
\bibinfo{author}{\bibfnamefont{V.~G.} \bibnamefont{Veselago}},
  \bibinfo{journal}{Soviet Physics Uspekhi} \textbf{\bibinfo{volume}{10}},
  \bibinfo{pages}{509} (\bibinfo{year}{1968}).

\bibitem[{\citenamefont{Pendry}(1996)}]{pendry1996}
\bibinfo{author}{\bibfnamefont{J.~B.} \bibnamefont{Pendry}},
  \bibinfo{journal}{Phys. Rev. Lett.} \textbf{\bibinfo{volume}{76}},
  \bibinfo{pages}{4773} (\bibinfo{year}{1996}).

\bibitem[{\citenamefont{Smith et~al.}(2000)\citenamefont{Smith, Padilla, Vier,
  Nemat-Nasser, and Schultz}}]{smith}
\bibinfo{author}{\bibfnamefont{D.~R.} \bibnamefont{Smith}},
  \bibinfo{author}{\bibfnamefont{W.~J.} \bibnamefont{Padilla}},
  \bibinfo{author}{\bibfnamefont{D.~C.} \bibnamefont{Vier}},
  \bibinfo{author}{\bibfnamefont{S.~C.} \bibnamefont{Nemat-Nasser}},
  \bibnamefont{and} \bibinfo{author}{\bibfnamefont{S.}~\bibnamefont{Schultz}},
  \bibinfo{journal}{Phys. Rev. Lett.} \textbf{\bibinfo{volume}{84}},
  \bibinfo{pages}{4184} (\bibinfo{year}{2000}).

\bibitem[{\citenamefont{Pendry}(2004)}]{pendry2004}
\bibinfo{author}{\bibfnamefont{J.~B.} \bibnamefont{Pendry}},
  \bibinfo{journal}{Contemporary Physics} \textbf{\bibinfo{volume}{45}},
  \bibinfo{pages}{191} (\bibinfo{year}{2004}).

\bibitem[{\citenamefont{Skaar}(2006{\natexlab{a}})}]{skaar06}
\bibinfo{author}{\bibfnamefont{J.}~\bibnamefont{Skaar}},
  \bibinfo{journal}{Phys. Rev. E} \textbf{\bibinfo{volume}{73}},
  \bibinfo{eid}{026605} (\bibinfo{year}{2006}{\natexlab{a}}).

\bibitem[{\citenamefont{Skaar}(2006{\natexlab{b}})}]{skaar06b}
\bibinfo{author}{\bibfnamefont{J.}~\bibnamefont{Skaar}}, \bibinfo{journal}{Opt.
  Lett.} \textbf{\bibinfo{volume}{31}}, \bibinfo{pages}{3372}
  (\bibinfo{year}{2006}{\natexlab{b}}).

\bibitem[{\citenamefont{Nistad and Skaar}(2008)}]{nistad08}
\bibinfo{author}{\bibfnamefont{B.}~\bibnamefont{Nistad}} \bibnamefont{and}
  \bibinfo{author}{\bibfnamefont{J.}~\bibnamefont{Skaar}},
  \bibinfo{journal}{Phys. Rev. E} \textbf{\bibinfo{volume}{78}},
  \bibinfo{pages}{036603} (\bibinfo{year}{2008}).

\bibitem[{\citenamefont{Pendry}(2000)}]{pendry2000}
\bibinfo{author}{\bibfnamefont{J.~B.} \bibnamefont{Pendry}},
  \bibinfo{journal}{Phys. Rev. Lett.} \textbf{\bibinfo{volume}{85}},
  \bibinfo{pages}{3966} (\bibinfo{year}{2000}).

\bibitem[{\citenamefont{Landau and Lifshitz}(1960)}]{landau_lifshitz_edcm}
\bibinfo{author}{\bibfnamefont{L.~D.} \bibnamefont{Landau}} \bibnamefont{and}
  \bibinfo{author}{\bibfnamefont{E.~M.} \bibnamefont{Lifshitz}},
  \emph{\bibinfo{title}{Electrodynamics of continuous media}}
  (\bibinfo{publisher}{Pergamon Press, New York and London, Chap. 9},
  \bibinfo{year}{1960}).

\bibitem[{\citenamefont{Ramakrishna et~al.}(2002)\citenamefont{Ramakrishna,
  Pendry, Schurig, Smith, and Schultz}}]{ramakrishna2002}
\bibinfo{author}{\bibfnamefont{S.~A.} \bibnamefont{Ramakrishna}},
  \bibinfo{author}{\bibfnamefont{J.~B.} \bibnamefont{Pendry}},
  \bibinfo{author}{\bibfnamefont{D.}~\bibnamefont{Schurig}},
  \bibinfo{author}{\bibfnamefont{D.~R.} \bibnamefont{Smith}}, \bibnamefont{and}
  \bibinfo{author}{\bibfnamefont{S.}~\bibnamefont{Schultz}},
  \bibinfo{journal}{J. Mod. Optics} \textbf{\bibinfo{volume}{49}},
  \bibinfo{pages}{1747} (\bibinfo{year}{2002}).

\bibitem[{\citenamefont{Lind-Johansen et~al.}(2009)\citenamefont{Lind-Johansen,
  Seip, and Skaar}}]{johansen}
\bibinfo{author}{\bibfnamefont{Ø.}~\bibnamefont{Lind-Johansen}},
  \bibinfo{author}{\bibfnamefont{K.}~\bibnamefont{Seip}}, \bibnamefont{and}
  \bibinfo{author}{\bibfnamefont{J.}~\bibnamefont{Skaar}}, \bibinfo{journal}{J.
  Math. Phys.} \textbf{\bibinfo{volume}{50}}, \bibinfo{pages}{012908}
  (\bibinfo{year}{2009}).

\end{thebibliography}

\end{document}